\documentclass[letterpaper,11pt]{nmeauth}
\usepackage{times}
\newcommand{\Fnsi}[1] {F_{\text{n,}#1}^{\ast}}
\newcommand{\Zetasi}[1] {\zeta^{\ast}_{#1}}
\newcommand{\dTb}[1]  {\Delta\overline{\boldsymbol{\theta}}^{#1}}
\newcommand{\Kb}[1]   {\overline{\mathbf{K}}^{#1}}
\newcommand{\dUp}[1]  {\Delta\overline{\mathbf{u}}^{#1}}

\newcommand{\Rp}[1]   {\overline{\mathbf{r}}^{#1}}
\newcommand{\dOmega}[1]{\Delta\overline{\omega}_{#1}}
\begin{document}
\NME{1}{6}{00}{00}{10}
\runningheads{M.\ R.\ Kuhn}
{J\"{a}ger Contact for {DEM} Simulations}
\title{Implementation of the J\"{a}ger contact model\\
       for discrete element simulations}
\author{Matthew R. Kuhn\corrauth\footnotemark[2]}
\address{University of Portland, 5000 N. Willamette Blvd.,
        Portland, OR, 97203~USA}
\corraddr{University of Portland, 5000 N. Willamette Blvd.,
          Portland, OR, 97203~USA,
          Tel.~(1)-503-943-7361,~Fax~(1)-503-943-7316}
\footnotetext[2]{E-mail: kuhn@up.edu}
\cgsn{National Science Foundation}{NEESR-936408}
\noreceived{}
\norevised{}
\noaccepted{}
\begin{abstract}
In three-dimensional discrete element (DEM) simulations, the particle motions 
within a granular assembly can produce
bewildering sequences of movements at the contacts
between particle pairs.
With frictional contacts, the relationship between contact movement and force
is non-linear and path-dependent,
requiring an efficient means of computing the forces 
and storing their histories.
By cleverly applying the principles of Cattaneo,
Mindlin, and Deresiewicz, J\"{a}ger~(2005) developed an efficient
approach for computing the full three-dimensional force between identical
elastic spheres that have undergone difficult movement sequences
(J.~J\"{a}ger, \emph{New Solutions in Contact Mechanics}, WIT Press,
Southampton, U.K.).
The paper presents a complete J\"{a}ger algorithm that can be incorporated
into DEM codes.
The paper also describes three
special provisions for DEM simulations:
(1) a method for handling particle pairs that undergo
complex tumbling and twirling motions in three-dimensions;
(2) a compact data structure for storing the loading history of the
many contacts in a large assembly; 
and (3) an approximation
of the J\"{a}ger algorithm that reduces memory demand.
The algorithm addresses contact translations 
between elastic spheres having identical properties,
but it does not
resolve the tractions produced by twisting or rolling motions.
A performance test demonstrates that the algorithm can be
applied in a DEM code with modest increases in computation
time but with more substantial increases in required storage.
\end{abstract}
\section{INTRODUCTION}
The discrete element (DEM) method is a computational approach for 
simulating granular materials by tracking the interactions of 
discrete, individual grains at their contacts.
Realistic simulations require a contact displacement-force 
model that is faithful 
to the physical grains that are being modeled.
In early implementations, simple linear springs were used to model
the normal and tangential forces between particles, with the tangential force
being limited by an abrupt frictional threshold
\cite{Cundall:1979a}.
Simulations that employ such simple models can yield qualitative agreement
with some granular phenomena, but they are inadequate
for the quantitative analysis of many situations.
For example, linear contact models are
incapable of accounting for the observed dependence of
bulk material stiffness and wave speeds upon the confining pressure,
which is an essential element in geotechnical problems.
Linear models also cannot account for the progressive and hysteretic
loss of stiffness in materials that are cyclically sheared with small
strain amplitudes, as in liquefaction problems.
\par
To account for these and other effects,
Hertz theory is now available in many DEM
codes to model the normal forces 
that develop between smooth elastic spheres
that are pressed together~\cite{Timoshenko:1951a}.
The corresponding
tangential shearing tractions that arise when particles shift 
in transverse, tangential directions
were first solved by Cattaneo and Mindlin, who reported the
progressive nature of frictional slip within the small circular
contact area between two spheres~\cite{Cattaneo:1938a,Mindlin:1949a}.
Such shearing tractions are history-dependent, and
the systematic analysis of loading histories usually begins with the work of 
Mindlin and Deresiewicz~\cite{Mindlin:1953a} 
who cataloged a canon of eleven basic histories
and derived analytic expressions for the resulting tractions.
This significant work was limited in two respects:
(1)~tangential loading and unloading were assumed 
to occur in a single tangential
direction with no provision for 
transverse tangential movements within the contact plane,
and (2)~the loading histories were limited 
to a few stages, with each additional stage bringing further
analytical difficulties.
\par
\begin{figure}
\centering
\includegraphics{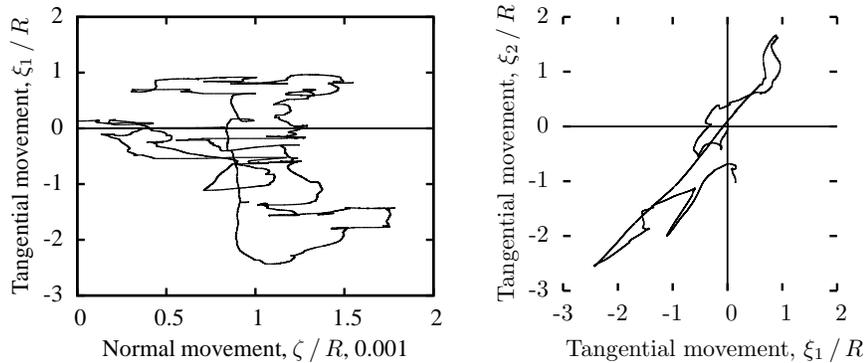}
\caption{\label{fig:DEMcontact}
         Normal and tangential movements of a typical DEM contact.}
\end{figure}

DEM simulations present far more complex situations.
Figure~\ref{fig:DEMcontact} shows the movement of a typical contact in a DEM 
simulation of slow (quasi-static) 
monotonic biaxial compression of a dense material.  
This particular contact, one of several thousand, became engaged 
after the peak stress state had been reached, and it 
remained engaged through an assembly strain of about 5\%, 
during which the normal movement $\zeta$ and the two components of 
tangential movement,
$\xi_{1}$ and $\xi_{2}$, were tracked.
The contact displays a bewildering sequence of loading and unloading
in all directions,
and this for a monotonic loading.
Although rapid collisional flows produce briefer
and simpler particle interactions, the behavior in
Figure~\ref{fig:DEMcontact}
is typical of slow flows of dense materials
in which the particles are in persistent contact across extended periods of 
bulk deformation.
Cyclic loading of dense assemblies will produce even more varied motions
at the contacts.
\par
Seridi and Dobry~\cite{Seridi:1984a} approached the 
first limitation by analyzing an
additional loading history in which tangential motion in
one direction was abruptly followed by 
an infinitesimal tangential movement in the 
perpendicular direction.
Placed in a continuum elasto-plasticity context,
they found that the result resembled a loading probe 
tangent to a yield surface.
With this insight,
they and their colleagues then resolved both of the
Mindlin-Deresiewicz limitations by developing a general
algorithm for the 
three-dimensional contact
displacement-force relation that
was analogous to incremental elasto-plasticity with kinematic 
hardening~\cite{Dobry:1991a}.
The algorithm approximates the force by using small increments
of movement and by tracking the loading history as a sequence of yield
cones.
Because of its computational demands, the method has
not been widely adopted.
%
\par
More recently, Vu-Quoc and Zhang~\cite{Vuquoc:1999a,Vuquoc:1999b}
developed another tangential force-displacement algorithm
that uses an incremental-stiffness form of the 
Mindlin-Deresiewicz model. 
They derived incremental stiffnesses
for four possible loading cases: combinations of tangential
and normal force increments that are each either increasing or decreasing.
The hysteretic behavior is captured by storing the tangential forces at which 
reversals in the loading direction occur (termed ``turning points'').
They also employed an improved form of a partially-latching
spring system for the normal forces~\cite{Walton:1986a,Vuquoc:1999a}.
This system models plastic deformation in the normal direction.
Zhang and Vu-Quoc~\cite{Zhang:2000a} demonstrated their contact
by modeling rapid, collisional flows of soybeans.
Most DEM codes now use a Hertzian normal force
combined with some incremental-stiffness form of the
Mindlin-Deresiewicz 
model (for example,~\cite{Thornton:1988a,Lin:1995a,Vuquoc:2000a}).
\par
In a remarkable series of papers and book, J\"{a}ger %
developed an elegant approach 
to the Cattaneo-Mindlin-Deresiewicz problem,
one that permits large and arbitrary
movements to be analyzed exactly and in whole
(notably~\cite{Jager:1993a,Jager:1996a,Jager:1998a,Jager:1999a,Jager:2003a,Jager:2005a}).
The method expresses an otherwise complex distribution
of shearing tractions
as a superposition of simple Cattaneo-Mindlin functions which provide
a compact means of chronicling the essential elements of quite
convoluted loading histories.
The author suggests that this method be named the
\emph{J\"{a}ger contact} or the \emph{J\"{a}ger algorithm}.
The paper presents a complete coding of the algorithm that is outlined in
the  J\"{a}ger text (\cite{Jager:2005a}, pp.~129--130).
The coding is intended for displacement-driven DEM simulations
and includes additions that assure its computational dependability
for arbitrary displacement histories.
\par
The J\"{a}ger algorithm computes the contact force for each step in
an arbitrary sequence of normal and tangential movements,
such as the movements in a series of DEM time steps.
The algorithm has several distinguishing characteristics:
\begin{itemize}
\item
Each normal-tangential movement can be of arbitrary size
without the need for incremental stiffnesses and without dividing
a large movement into smaller, incremental sub-movements.
In this sense, the J\"{a}ger algorithm reproduces 
the closed-form solutions of 
Mindlin and Deresiewicz (such as equations~7 and~14 in~\cite{Mindlin:1953a})
that give the accumulated contact
displacement in terms of the full contact force.
(Mindlin-Deresiewicz also present instantaneous, incremental
stiffnesses and compliances, but these are not part of the
J\"{a}ger method.)
The method only requires that the total normal movement
is small when compared with the particle radius 
and that each movement is monotonic, such that motion
advances in a continuous and proportional manner within
each movement.
\item
The algorithm is inherently three-dimensional, 
delivering the final three-dimensional force for
an arbitrary sequence of normal and 2-vector 
tangential movements~\cite{Jager:2003a,Jager:2005a}.
\item
The algorithm includes a rigorous means of identifying load
reversals (turning-points) in a full three-dimensional setting,
and it incorporates the effect of such reversals upon subsequent
loading steps.
\item
As shown by Mindlin and Deresiewicz~\cite{Mindlin:1953a},
certain combinations of tangential and normal movements will
suppress slip within the circular contact area between elastic spheres.
J\"{a}ger refers to such movements as producing a stick condition, but
such movements are also called elastic 
or non-simple~\cite{Vuquoc:1999a}.
Even though
slip is suppressed, such movements will alter the distribution
of contact tractions and, hence, will affect the onset and extent
of slip in future movements.
Mindlin and Deresiewicz~\cite{Mindlin:1953a}
do not address non-simple loading sequences in which
a general elastic movement is followed by a non-elastic movement.
J\"{a}ger derived a solution to this problem by following the
tractions in the current and past slip zones of a contact area
and the manner in which the Cattaneo-Mindlin tractions must be superposed
(see~\cite{Jager:1998a,Jager:1999a} for succinct derivations).
\par
\quad
Because both elastic and non-elastic movements can affect future slip,
the J\"{a}ger algorithm thoroughly chronicles any
changes in the directions of either elastic or non-elastic movements,
including those at turning-points.
The algorithm then accounts for the effect that past elastic and non-elastic
movements will have upon the current and future forces.
In short, the J\"{a}ger algorithm exactly computes the contact force
for non-simple histories, and it does so without the use of
incremental stiffnesses or compliances.
\end{itemize}
\par
Besides presenting the coding for the J\"{a}ger algorithm,
the paper also describes three special provisions that
apply to DEM simulations:
(1) a method that is necessary for applying
the J\"{a}ger algorithm to particle pairs that undergo the 
twirling and tumbling motions that can be expected in three-dimensional
DEM simulations, i.e. motions that will also twirl and rotate
the entire contact force;
(2) an approximation
of the J\"{a}ger contact that reduces memory demand;
and (3) a compact data structure for storing the loading histories of the
many contacts in a large assembly.
\par
The algorithm (and the paper) is not without limitations, 
as it only concerns contact translations and
does not account for tractions produced by the twisting (torsional) rotations
of two particles about their contact normal direction, nor does
it account for tractions produced by the rolling between particles
(rolling friction).
The paper's underlying material model is one of isotropic 
linear elasticity for two spheres having equal elastic properties, 
an assumption carried over from the original
works of Hertz, Cattaneo, Mindlin, and Deresiewicz.
The method does not account for contact adhesion or for
plasticity, visco-elasticity, or fracture and breaking 
of the spheres.
Investigators have recently modeled the contact problem as
an elasto-plastic process~\cite{Walton:1986a,Thornton:1997b}.
Vu-Quoc and Zhang modeled the normal force-displacement
relationship of elasto-plastic 
spheres~\cite{Vuquoc:1999c,VuQuoc:2000b}
and, more recently, developed
an algorithm for the tangential and normal 
interactions of elasto-plastic spheres~\cite{VuQuoc:2004a,Zhang:2007a}.
Experimental validation of this advanced normal force model
is provided in~\cite{Plantard:2005a} for polymer spheres.
The paper assumes that the particles are spherical at their
contact, although J\"{a}ger's theory encompasses 
other shapes~\cite{Jager:2005a}.
\par
The next section describes the J\"{a}ger contact and
provides detailed pseudo-code of its algorithm.
Section~\ref{sec:implementation} 
supplies three additional provisions (listed above) that
were not part of J\"{a}ger's original method
but are required for an efficient DEM implementation.
The algorithm's performance in a large DEM simulation is described in
Section~\ref{sec:perform}, 
which also includes a simple example that demonstrates the
exact correspondence of the J\"{a}ger and Mindlin-Deresiewicz results
and compares the J\"{a}ger solution with an incremental-stiffness solution.
\section{J\"{A}GER ALGORITHM}
\begin{figure}[tb]
\centering
\includegraphics[scale=0.9]{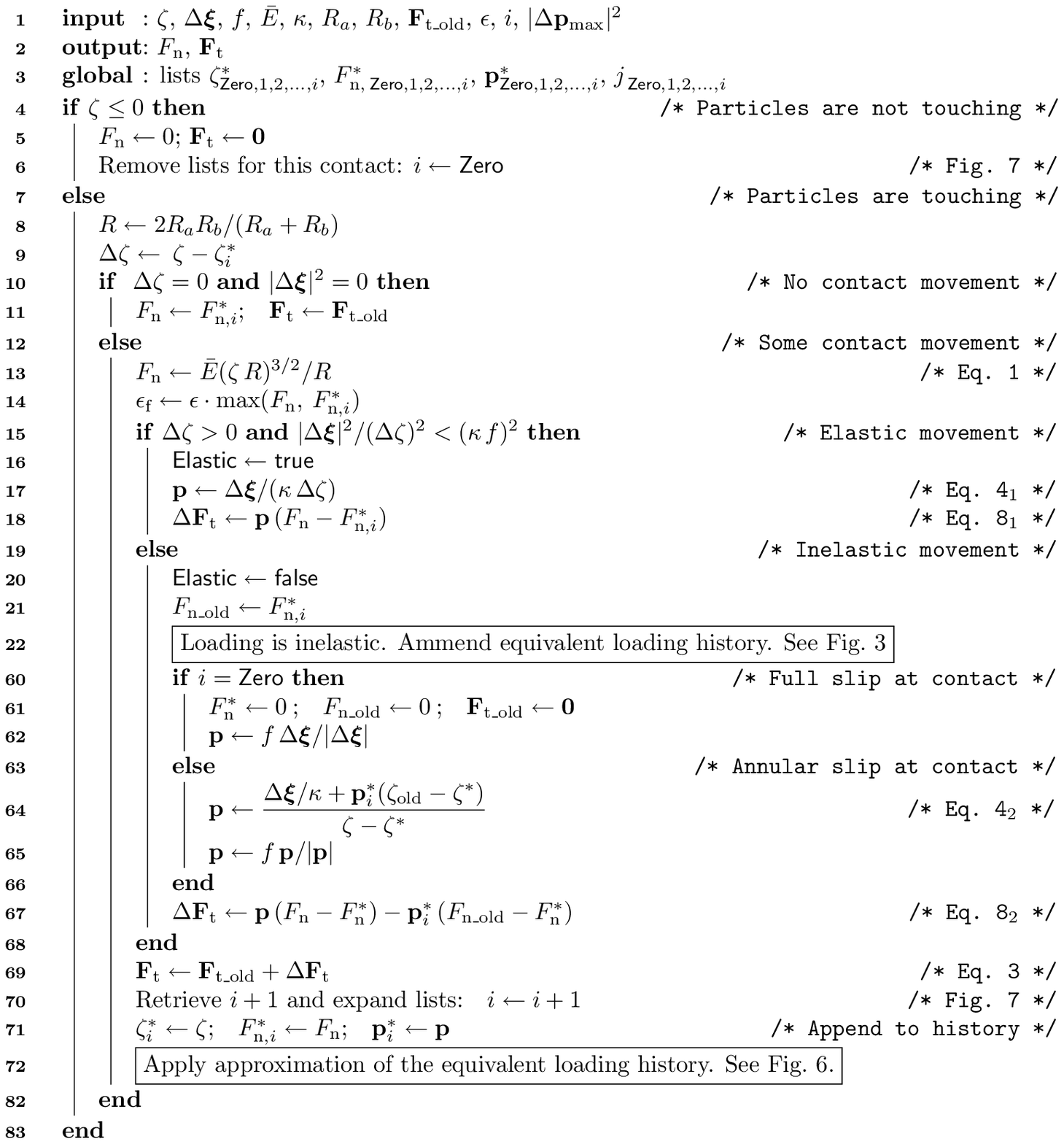}
\caption{\label{fig:Jager1}Algorithm of the J\"{a}ger contact.}
\end{figure}

\begin{figure}[htb]
\centering
\includegraphics[scale=0.85]{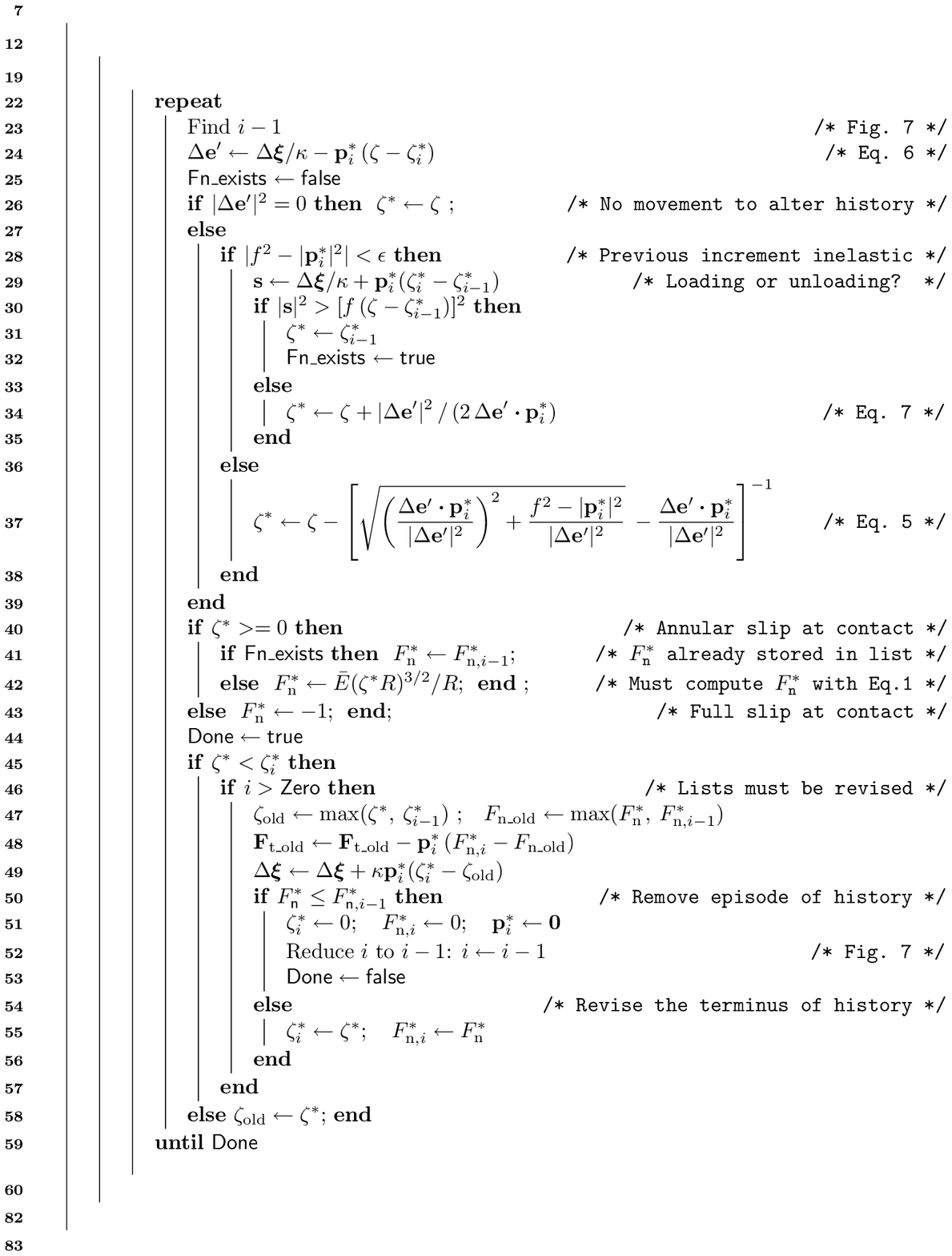}
\caption{\label{fig:Jager2}Lines 22--59 within the J\"{a}ger algorithm of 
Figure~\ref{fig:Jager1}.}
\end{figure}

We consider two identical isotropic-elastic spheres that undergo an arbitrary
sequence of finite, perhaps large, translations.
Contact forces $F_{\text{n}}$ and $\mathbf{F}_{\text{t}}$ are the compressive normal
force and the 2-vector tangential force.
Displacements $2\zeta$ and $2\boldsymbol{\xi}$ 
are the cumulative normal approach (overlap)
and the 2-vector tangential shift of the particles' centers.
Figures~\ref{fig:Jager1} and~\ref{fig:Jager2}
present the J\"{a}ger algorithm that is outlined in 
\cite{Jager:1996a} and~\cite{Jager:2005a}, \S7.2.
The two figures also include the author's additions that improve 
computational dependability and efficiency.
The figures present a procedure (function) that would be called
within each time step and for each contact within
a DEM assembly.
\par
For the given input (line~1) the algorithm returns
the contact's normal and 2-vector tangential forces, $F_{\text{n}}$ 
and $\mathbf{F}_{\text{t}}$ (line~2).
The input includes 
the contact half-overlap (indentation) $\zeta$;
the 2-vector tangential half-shift movement $\Delta\boldsymbol{\xi}$;
the friction coefficient $f$;
the modulus $\overline{E}=8G/(3(1-\nu))$,
which depends on the particles' shear modulus $G$ and Poisson ratio $\nu$;
the ratio $\kappa$ of elastic normal and tangential stiffnesses,
$\kappa = (2-\nu)/(2(1-\nu))$;
the particles' radii of curvature, $R_{A}$ and $R_{B}$; 
the previous tangential force $\mathbf{F}_{\text{t\_old}}$;
a small tolerance parameter $\epsilon$;
a pointer $i$ to the top of an equivalent load history stack, 
described in Section~\ref{sec:lists};
and a parameter $|\Delta\mathbf{p}_{\text{max}} |^{2}$ 
to reduce memory demand (Section~\ref{sec:approx}).
Lists that store the equivalent load history are included
as both input and output (line~3).
The contents of these lists 
(i.e., $\zeta^{\ast}_{\ldots}$, $F^{\ast}_{\text{n,}\ldots}$,
$\mathbf{p}^{\ast}_{\ldots}$, and $j_{\ldots}$)
are described later.
\par
The Hertz normal force only depends on the 
indentation $\zeta$~\cite{Timoshenko:1951a}:
\begin{equation}
F_{\text{n}} = \overline{E} \sqrt{\zeta^{3} R} \label{eq:Fn}
\end{equation}
where the modulus $\overline{E}=8G/(3(1-\nu))$ (line~13).
For spheres of different radii, an approximate average radius is 
adopted on line~8~\cite{Johnson:1985a}.
\par
In the work of Mindlin and Deresiewicz~\cite{Mindlin:1953a},
most equations for tractions, compliances, and displacements refer
to the radii of various contact and stick areas.
For example, the full contact area has radius
\begin{equation}
a = \left( \frac{F_{\text{n}}R}{\overline{E}} \right)^{1/3} \label{eq:a}
\end{equation}
and they gave the radii 
of stick areas the symbols $b$, $c^{\ast}$, etc.
Although these radii could be used within the J\"{a}ger algorithm,
it is more convenient to use $\zeta$ and $F_{\text{n}}$ as simpler
proxies.  
In this regard, equations~(\ref{eq:Fn}) and~(\ref{eq:a})
can be used to shift between the results in the paper and those
of Mindlin and Deresiewicz.
\par
Once the particles touch, 
the tangential force $\mathbf{F}_{\text{t}}$ will depend upon the DEM history of the finite
movements (translations) between time steps, the
$\Delta\zeta$ and $\Delta\boldsymbol{\xi}$, that
lead to the contact's current position $(\zeta,\boldsymbol{\xi})$.
The current force $\mathbf{F}_{\text{t}}$ is produced by a force
increment $\Delta\mathbf{F}_{\text{t}}$ 
relative to the previous force (i.e., from the previous 
time step):
\begin{equation}
\mathbf{F}_{\text{t}} = \mathbf{F}_{\text{t\_old}} + \Delta\mathbf{F}_{\text{t}}
\label{eq:newFt}
\end{equation}
(line~71).  
The tangential increment $\Delta\mathbf{F}_{\text{t}}$ 
is entirely elastic, producing no slip, when the current movement
$(\Delta\zeta ,\Delta\boldsymbol{\xi} )$ satisfies the 
following two conditions:
$\Delta\zeta >0$ and $|\Delta\boldsymbol{\xi} / (\kappa \Delta\zeta )|<f$
(line~15, as demonstrated in~\cite{Mindlin:1953a}, 
\S14, 15, and~18).
When either condition is violated, the current movement will produce
frictional slip within an outer annular ring of the circular contact area
between the spheres~\cite{Mindlin:1953a}.
\par
The current movement will not only affect the current increment of tangential
force but can affect future increments as well.
A journal of the contact's past loading, in the form of
an \emph{equivalent load history} is maintained, so that the
current and future loads will be reconciled with this history.
The contact's equivalent ($\ast$-star) load
history is a compact recording of an equivalent sequence of loading steps
that would lead to the previous contact force
$\mathbf{F}_{\text{t,old}}$ and the cumulative displacement
$\boldsymbol{\xi}_{\text{old}}$.
The equivalent load history 
is stored as lists of the normal indentations
and the corresponding normal forces~--- 
the lists $\zeta^{\ast}_{0}$, $\zeta^{\ast}_{1},\;\ldots$ $\zeta^{\ast}_{i}$
and $\Fnsi{0}$, $\Fnsi{1},\;\ldots$ $\Fnsi{i}$~---
along with a list of 2-vector directions of 
tangential force, $\mathbf{p}^{\ast}_{0}$, $\mathbf{p}^{\ast}_{1},\;\ldots$ $\mathbf{p}^{\ast}_{i}$.
Each $\zeta^{\ast}$-$F_{\text{n}}^{\ast}$ pair designates the apex of
a yield cone in displacement-space and force-space.
In another sense, the $\zeta^{\ast}$ and $F_{\text{n}}^{\ast}$
are proxies for the radii of past slip areas, as in
equation~(\ref{eq:a}) and in reference~\cite{Mindlin:1953a}.
This ``$\ast$'' data gives information about the points of load reversals
(turning-points) as well as information from past elastic movements
that will affect the onset and the extent of future frictional slip.
Upon entering the procedure, the history's index will range from
$0$ to $i$.
The zeroth point is permanently initialized to the 
unloaded condition: $\zeta^{\ast}_{0}=\Fnsi{0}=0$ and $\mathbf{p}^{\ast}_{0}=\mathbf{0}$.
Within the procedure, the lists' lengths can be increased by~1, 
remain the same,
or be reduced to as small as~2 (lines~52 and~70).
Before leaving the procedure,
the most recently computed values of $\zeta$, $F_{\text{n}}$, and $\mathbf{p}$ are
appended to the equivalent history (lines~70--71).
Because these lists are more compactly stored as linked lists, the
indices 0,1,$\ldots i$ are, in reality, pointers to
locations within such linked lists (Section~\ref{sec:lists}).
\par
\begin{figure}
\centering
\includegraphics{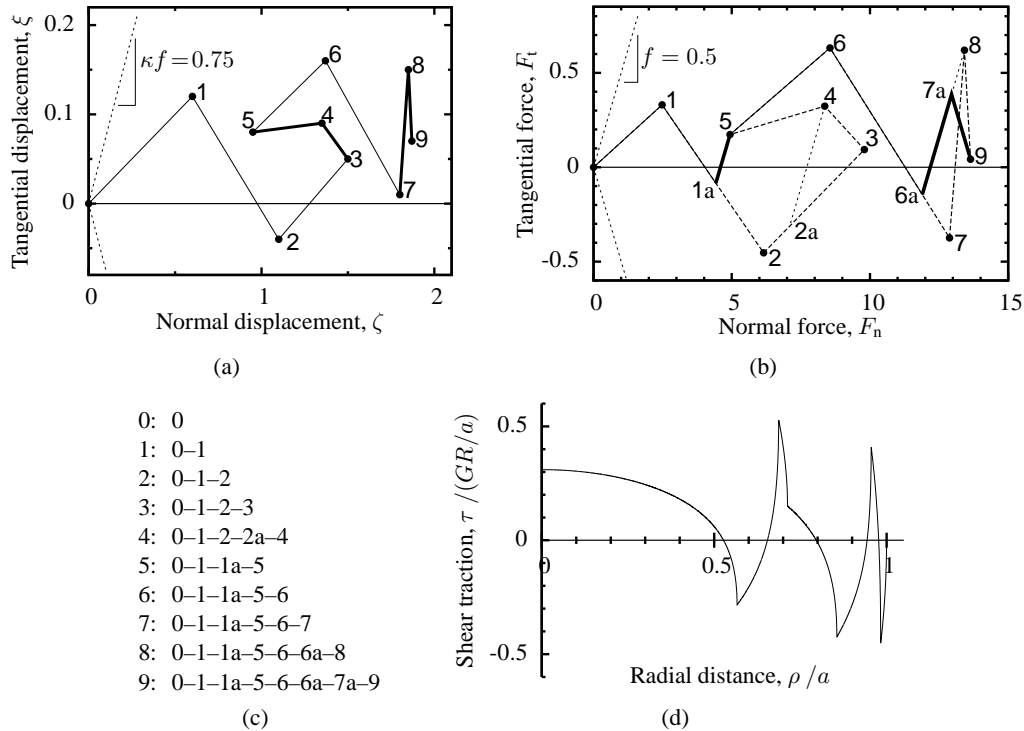}
\caption{\label{fig:JagerPath}
  Example sequence of (a) nine contact movements, (b) contact forces,
  (c) equivalent load histories, and (d) shearing traction.
  In (b), the solid line is the final equivalent
  load history; dashed lines are intermediate histories.
  In (d), final shearing tractions 
  are across the radius $a$ of the circular contact
  area at the end of the step~9
  ($f=0.5$, $\kappa f=0.75$, $G=1$, $\nu=0.5$,
  and $R=1$).}
\end{figure}

The concept of an equivalent load history is illustrated in
Figure~\ref{fig:JagerPath}.
In this simple two-dimensional example, the tangential movements are
colinear, so only single components of the two-vectors 
$\boldsymbol{\xi}$ and $\mathbf{F}_{\text{t}}$
are relevant.
The first part, Figure~\ref{fig:JagerPath}a, 
shows an example sequence of nine 
movements between two spheres having friction
coefficient $f=0.5$ and $\kappa f=0.75$,
where $\kappa = (2-\nu)/(2(1-\nu))$.
The particles first touch at stage ``0'' and then undergo a
sequence of nine proportional (straight) normal-tangential movements:
$(\Delta\zeta_{1} , \Delta\xi_{1})$, 
\mbox{$(\Delta\zeta_{2} , \Delta\xi_{2})$,$\ldots$,}
$(\Delta\zeta_{9} , \Delta\xi_{9})$,
These successive movements produce the corresponding
sequence of forces in the second part, Figure~\ref{fig:JagerPath}b,
labeled as the points 1,~2,~$\ldots$,~9.
The solid line in Figure~\ref{fig:JagerPath}b gives the final equivalent
load history at the end of step~9.
This equivalent history \mbox{(path 0--1--1a--$\ldots$--7a--9)} would
produce the same final displacement 
($\zeta_{9}$, $\boldsymbol{\xi}_{9}$), 
the same contact force ($F_{\text{n,9}}$, $\mathbf{F}_{\text{t,9}}$), 
and the same tractions as the original
nine steps \mbox{(0--1--2--$\ldots$--8--9)}.
The dashed lines in Figure\ref{fig:JagerPath}b
trace the intermediate equivalent histories that
would lead to the cumulative intermediate forces $\mathbf{F}_{\text{t,1}}$,
$\mathbf{F}_{\text{t,2}}$, etc.
These intermediate (dashed) segments were eventually eliminated within
the equivalent history that leads to the 
final force $\mathbf{F}_{\text{t,9}}$.
For example, the sequence 0--1--1a--5 will produce the same cumulative
displacement ($\zeta_{5}$, $\boldsymbol{\xi}_{5}$)
as the sequence 0--1--2--3--4--5, so only the
truncated (solid) sequence is stored as the equivalent load history 
at step~5.
Although five of the nine movements are elastic
(1, 2, 3, 6, and~7), four have slopes
$\Delta\xi / \Delta\zeta$ that violate the elastic condition
and thus produce annular slips (4, 5, 8, and~9,
shown as the darker lines in Figure~\ref{fig:JagerPath}a).
The evolution of the equivalent load history is listed
in Figure~\ref{fig:JagerPath}c, with all segments constrained to
a slope $\mathbf{p}$ no steeper than $\pm f$.
The four slip movements result in equivalent path segments that have
a slope magnitude $|\Delta\mathbf{F}_{\text{t}} / \Delta F_{\text{n}}|=|\mathbf{p}|=f$:
the segments 2a--4, 1a--5, 6a--8, and 7a--9.
Along the final (solid) equivalent load history in
Figure~\ref{fig:JagerPath}b,
three of the segments are shown with heavier solid lines, as they
are non-elastic (slip) loadings that lead to the final
force $\mathbf{F}_{\text{t,9}}$.
The remaining (thinner) solid lines are elastic segments, yet they
must be stored since they
can also affect the future onset and extent of frictional slip
(i.e., the loading sequence is non-simple).
Figure~\ref{fig:JagerPath}d shows the final distribution of shear traction
across the circular contact area of radius $a$, revealing
remnants of the seven equivalent force steps as seven superposed
Cattaneo-Mindlin functions.
If needed, the traction distribution can be constructed with
the J\"{a}ger algorithm using principles given in 
Section~\ref{section:compare}.
\par
Although
the equivalent load sequence of the two-dimensional
nine-step example in Figure~\ref{fig:JagerPath} is somewhat complex, 
the full three-dimensional
sequences in DEM simulations are far more tortuous: 
instead of a length of 7, the lengths are often greater than 100
(see Figure~\ref{fig:DEMcontact}).
\par
We now return to the J\"{a}ger algorithm of
Figures~\ref{fig:Jager1} and~\ref{fig:Jager2}.
A finite movement $(\Delta\zeta ,\Delta\boldsymbol{\xi} )$ 
advances the normal indentation,
$\zeta = \zeta_{\text{old}} + \Delta\zeta$, 
and the new normal force $F_{\text{n}}$ is
computed with equation~(\ref{eq:Fn}) (line~13).
J\"{a}ger showed that the tangential movement 
$\Delta\boldsymbol{\xi}$ can be expressed
in terms of three indentations:
the current indentation $\zeta$, the previous indentation
$\zeta_{\text{old}}$ (stored as $\Zetasi{i}$), and an effective
(equivalent) indentation $\zeta^{\ast}$:
\begin{equation}
\Delta\boldsymbol{\xi} = \begin{cases}
       \kappa\mathbf{p}(\zeta - \zeta_{\text{old}} )\:, & \Delta\zeta > 0 \text{ and } 
                   |\Delta\boldsymbol{\xi} /(\kappa\Delta\zeta )|<f \\
       \kappa\mathbf{p}(\zeta - \zeta^{\ast} ) - \kappa\mathbf{p}^{\ast}_{i} (\zeta_{\text{old}} - \zeta^{\ast} ) 
             \text{ and } |\mathbf{p}|=f\:, &
             \text{otherwise}
       \end{cases}
\label{eq:dXi}
\end{equation}
Each indentation corresponds to a normal force
($F_{\text{n}}$, $F_{\text{n,old}}$, and $F_{\text{n}}^{\ast}$) 
and a contact radius through equations~(\ref{eq:Fn}) and~(\ref{eq:a}).
The first case in equation~(\ref{eq:dXi})
is for an elastic movement, and~(\ref{eq:dXi}$_{1}$)
can be rearranged to
give the direction of tangential force, 
$\mathbf{p} = \Delta\mathbf{F}_{\text{t}} / \Delta F_{\text{n}} = \Delta\boldsymbol{\xi} / (\kappa \Delta\zeta )$ (line~17).
\par
The second case~(\ref{eq:dXi}$_{2}$) 
applies when the movement produces annular slip
within the contact area.
In this case, the norm $|\mathbf{p}|$ is capped at the friction coefficient
$f$, and direction $\mathbf{p}$ is aligned with the tangential movement, 
$\mathbf{p} = f\,\Delta\boldsymbol{\xi}/|\Delta\boldsymbol{\xi} |$ 
(lines~62 and~65).
When used in a displacement-driven DEM algorithm, the second case serves as
the consistency condition for establishing the most recent point
$\zeta^{\ast}$ of the equivalent history, which is used later to
find $\Delta\mathbf{F}_{\text{t}}$.
(This situation is similar to using a consistency principle to locate
the back-stress in conventional elasto-plasticity with kinematic
hardening.)
When the previous movement is elastic (with $|\mathbf{p}^{\ast}_{i}| < f$
but $|\mathbf{p}| = f$),
equation~(\ref{eq:dXi}$_{2}$) is rearranged as
\begin{equation}
\zeta^{\ast} = \zeta -
\left[
  \sqrt{ \left( \frac{\Delta\mathbf{e}^{\prime} \boldsymbol{\cdot}
  \mathbf{p}^{\ast}_{i}}{|\Delta\mathbf{e}^{\prime} |^{2}} \right)^{2} 
  + \frac{f^{2} \:-\: 
  |\mathbf{p}^{\ast}_{i}|^{2}}{|\Delta\mathbf{e}^{\prime} |^{2}}\,}
-
\frac{\Delta\mathbf{e}^{\prime} 
  \boldsymbol{\cdot} \mathbf{p}^{\ast}_{i}}{|\Delta\mathbf{e}^{\prime} |^{2}}
\right]^{-1}
\label{eq:Zetas}
\end{equation}
where
\begin{equation}
\Delta\mathbf{e}^{\prime} = \frac{1}{\kappa} \Delta\boldsymbol{\xi} 
      - \mathbf{p}^{\ast}_{i} (\zeta - \zeta_{\text{old}} )
\label{eq:deprime}
\end{equation}
as in lines~24 and~37 (see~\cite{Jager:2005a}, p.~128).
On the other hand, when the previous movement also produces
slip (with $|\mathbf{p}^{\ast}_{i}| = f$),
we must distinguish between current movements 
$\Delta\boldsymbol{\xi}$ that produce
continued loading~--- thus expanding the current yield cone with apex at
$\Zetasi{i-1}$~--- and movements that produce unloading within
the cone 
(in Figure~\ref{fig:JagerPath}b, 
the former is represented by steps~4, 5, and~8; the latter by step~9).
Loading occurs when the conditions on lines~29--30
are satisfied.
When the movement produces unloading, equation~(\ref{eq:dXi}$_{2}$)
is rearranged as
\begin{equation}
\zeta^{\ast} = \zeta + \frac{1}{2}\frac{|\Delta\mathbf{e}^{\prime}|^{2}}
       {\,\Delta\mathbf{e}^{\prime}\boldsymbol{\cdot}\mathbf{p}^{\ast}_{i}}
\label{eq:Zetas2}
\end{equation}
as on line~34.
\par
The equivalent force $F_{\text{n}}^{\ast}$ that corresponds to 
$\zeta^{\ast}$ is computed with
equation~(\ref{eq:Fn}) (lines~40--42). 
Once $\zeta^{\ast}$ and $F_{\text{n}}^{\ast}$ are determined, 
J\"{a}ger's force-space complement of
(\ref{eq:dXi}) is used to
compute the tangential force increment $\Delta\mathbf{F}_{\text{t}}$:
\begin{equation}
\Delta\mathbf{F}_{\text{t}} = \begin{cases}
         \mathbf{p} (F_{\text{n}} - F_{\text{n\_old}} ) \\
         \mathbf{p} (F_{\text{n}} - F_{\text{n}}^{\ast} ) 
          - \mathbf{p}^{\ast}_{i} (F_{\text{n\_old}} - F_{\text{n}}^{\ast} )
       \end{cases}
\label{eq:dFt}
\end{equation}
as on lines~18 and~67.
\par
The sequence $\Zetasi{0},\;\Zetasi{1},\;\ldots\;\Zetasi{i}$
must be monotonically increasing
(in an elasto-plasticity setting, this requirement obviates
the intersection of yield cones).
When slip occurs, equations~(\ref{eq:Zetas}) 
and~(\ref{eq:Zetas2}) will give a value
$\zeta^{\ast} \le \Zetasi{i} (= \zeta_{\text{old}} )$,
and we must amend the load history: 
(1) replacing $\Zetasi{i}$ and $\Fnsi{i}$ with
$\zeta^{\ast}$ and $F_{\text{n}}^{\ast}$, 
(2) shifting $\zeta_{\text{old}}$ and $F_{\text{n\_old}}$ to $\zeta^{\ast}$ and $F_{\text{n}}^{\ast}$,
and (3) altering $\mathbf{F}_{\text{t\_old}}$ and $\Delta\boldsymbol{\xi}$ so that they originate
from $F_{\text{n\_old}}$ and $\zeta_{\text{old}}$
(lines~49--51 and~57).
In Figure~\ref{fig:JagerPath},
these operations would replace the sequence 0--1--2--3--4 with
its equivalent sequence 0--1--2--2a--4, and they
replace 0--1a--5--6--7--8 with 0--1a--5--6--6a--8.
In some cases, $\zeta^{\ast} \le \Zetasi{i-1}$, so that the intermediate
episode $\Zetasi{i}$ must be eliminated in a repetitive manner 
(lines~22 and~50--53):
in this way, the sequence 0--1--2--2a--4--5 is replaced with 0--1--1a--5.
\par
The algorithm in Figures~\ref{fig:Jager1} and~\ref{fig:Jager2}
is based on the one outlined by J\"{a}ger~\cite{Jager:2005a}.
The lines~28--35 are added for the special case
of two successive time steps that produce slip, as in 
equation~(\ref{eq:Zetas2})
(since equation~\ref{eq:Zetas} would otherwise involve division by zero).
The small tolerance $\epsilon$ is added in
line~28 to avoid computational problems when comparing
slopes of force
(the author has used an $\epsilon = 1\times 10^{-10}$ in his trials).
Lines~25 and~41 reduce the computational
demands of applying equation~(\ref{eq:Fn}) by retrieving previously computed
values of $F_{\text{n}}^{\ast}$ from memory.
Three additional provisions are described in Section~\ref{sec:implementation}.
\par
The algorithm has been implemented by the author in the OVAL DEM code
to simulate the quasi-static loading of large 
assemblies of particles~\cite{Kuhn:2002b}.
At the time of writing, the algorithm has been robust through about 
fifty billion procedure calls.
The algorithm's performance is also described in Section~\ref{sec:perform}.
\section{IMPLEMENTATION DETAILS}\label{sec:implementation}
\subsection{An approximate load history}\label{sec:approx}
\begin{figure}
\figcap{4.50in}
\centering
\includegraphics[scale=0.90]{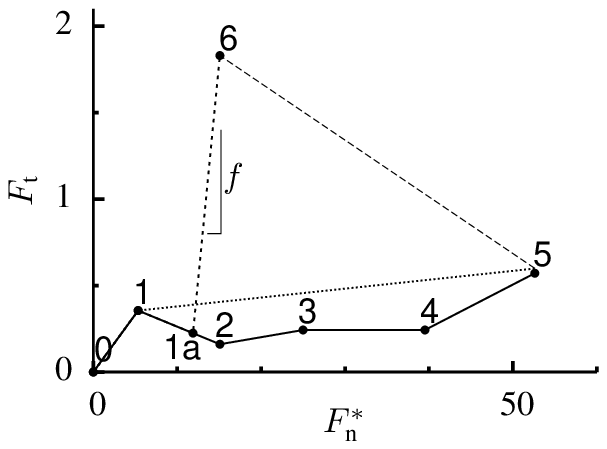}
\caption{\label{fig:Jager_approx}
         A memory-saving approximation of the equivalent loading history 
         ($f=0.5$, $|\Delta\mathbf{p}_{\text{max}} |=0.05$, $\kappa f=0.75$, $G=1$, $\nu=0.5$,
          and $R=1$).}
\end{figure}

\begin{figure}[tb]
\centering
\includegraphics[scale=0.9]{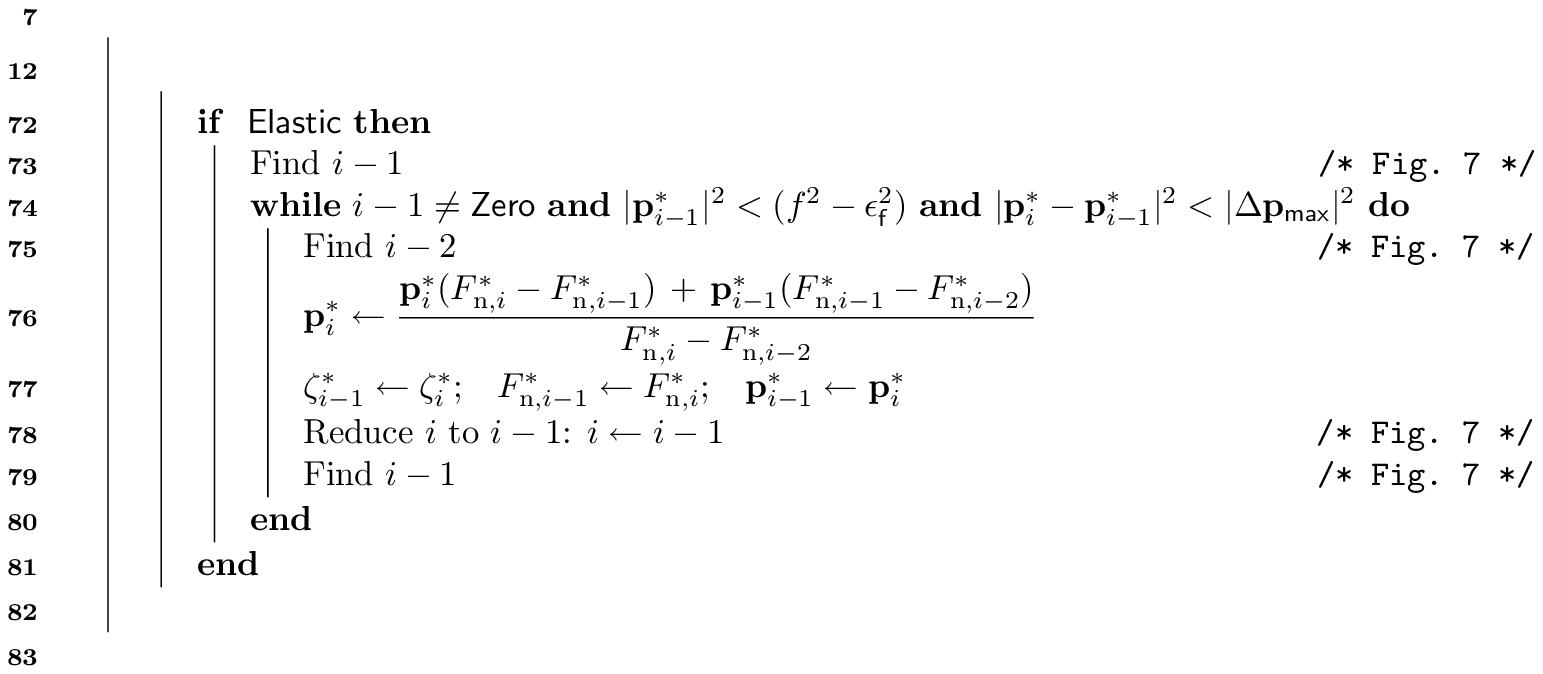}
\caption{\label{fig:Jager3}
         Lines 72--81 within the algorithm of Figure~2 to reduce memory demand.}
\end{figure}

The algorithm requires sufficient memory to store the 
three lists $\zeta^{\ast}_{0,1,\ldots,i}$,
$F^{\ast}_{\text{n},0,1,\ldots,i}$, and
$\mathbf{p}^{\ast}_{0,1,\ldots,i}$.
The demand for memory is particularly acute during gradual or
quiescent DEM simulations in which contacts remain elastic across
many time steps, causing these lists to grow with each passing step.
In Figure~\ref{fig:Jager_approx}, the list for this contact grows to length~6
during the gradual sequence of elastic steps~0 to~5; whereas the
single inelastic step~6 suddenly reduces the list to length~4
(0--1--1a--6) and releases memory that can be used elsewhere.
As an approximation, 
the lists can be reduced artificially by systematically combining several
similar elastic segments of the equivalent loading history.
This approximation is achieved by inserting the code of
Figure~\ref{fig:Jager3} into that of Figure~\ref{fig:Jager1},
lines~72--81.
When the $\mathbf{p}^{\ast}$ slopes 
of two successive elastic segments differ by less
than a prescribed small amount $|\Delta\mathbf{p}_{\text{max}}|$
(an input parameter with range 0 to $2f$), 
the approximation combines the two segments into one, 
thus releasing memory 
(lines 76--78, in which an average slope $\mathbf{p}_{i}^{\ast}$ 
is computed and the intermediate point, $i-1$, is removed).
This process is applied repeatedly to combine successive groups of segments
(line~74).
The approximation algorithm is illustrated in 
Figure~\ref{fig:Jager_approx} in which the four
segments 1--2, 2--3, 3--4, and 4--5 have a similar
slope, with $|\mathbf{p}^{\ast}_{i} - \mathbf{p}^{\ast}_{i-1}| \le |\Delta\mathbf{p}_{\text{max}} |$, and are combined into
the single (stippled) segment 1--5.
Although this modification introduces a small 
error into the history
that can affect future inelastic loading steps,
the memory savings are considerable.  
The error in approximating $\mathbf{F}_{\text{t,}6}$ 
is about 3\% for the example in Figure~\ref{fig:Jager_approx},
whereas the history-lists
can be reduced by over 30 percent when
the ratio $|\Delta\mathbf{p}_{\text{max}} |/f$ is as small as 0.10.
\subsection{Contact frame rotations}
In the previous sections, the two particles were assumed only to translate,
thus producing the contact movements $\Delta\zeta$ and $\Delta\boldsymbol{\xi}$.
In a more general setting, two contacting particles
can both move and rotate in several modes, and the DEM code must extract
the single mode of contact translation before the J\"{a}ger algorithm
can be applied.
The four modes of movement between two particles are a contact translation;
a rotational contact twisting; a rolling of the particles at their contact;
and a rigid rotation of the pair (see~\cite{Kuhn:2004k}).
Although twisting and rolling are objective motions that can alter the
traction within the circular contact area, they are not considered
in the paper~--- only contact translations are assumed to produce the 
deformations that give rise to $F_{\text{n}}$ and $\mathbf{F}_{\text{t}}$.
Rigid rotations must be considered, however, as they will rotate
a contact's coordinate frame.
In an extreme case, two particles can twirl and tumble as a rigid
(glued) pair while producing no contact translations at all, even as their
contact force 
is twirled and reoriented within the global (assembly) frame.
\par
The contact translation $\Delta\overline{\mathbf{u}}^{\text{def}}$ of particle $q$ 
relative to particle $p$ is
\begin{equation}
\Delta\overline{\mathbf{u}}^{\text{def}} = \dUp{q} - \dUp{p} 
         + \left( \dTb{q}\times\Rp{q} - \dTb{p}\times\Rp{p} \right)
\label{eq:dudef}
\end{equation}
where $\dUp{p}$, $\dUp{q}$, $\dTb{p}$, and $\dTb{p}$
are the particle translations and rotations,
and $\Rp{p}$ and $\Rp{q}$ are the vectors from the centers 
of particles $p$ and $q$ to the contact.
In this equation, all overbar quantities
(for example, $\dUp{p}$) designate vectors
viewed within the global (assembly) coordinate frame.
This view will differ from that within the local contact frame of
the particle pair (no overbar),
and it is within this local frame that the J\"{a}ger algorithm operates.
The translation vector can be split into normal and tangential parts,
\begin{align}
\Delta\zeta &= -\frac{1}{2}\left( \Delta\overline{\mathbf{u}}^{\text{def}} \boldsymbol{\cdot} \overline{\mathbf{n}}^{p} \right)
  \label{eq:dzeta} \\
\Delta\overline{\boldsymbol{\xi}}  &= \frac{1}{2} \left( \Delta\overline{\mathbf{u}}^{\text{def}} + 2\,\Delta\zeta\,\overline{\mathbf{n}}^{p} \right)
  \label{eq:dXib}
\end{align}
where $\overline{\mathbf{n}}^{p}$ is the outward unit normal vector of $p$ at the contact.
\par
In a DEM code, we compute the $\Delta\overline{\boldsymbol{\xi}}$ movement in the global frame
(equation~\ref{eq:dXib});
transform (rotate) the global $\Delta\overline{\boldsymbol{\xi}}$ and $\overline{\mathbf{F}}_{\text{t\_old}}$ vectors into the local
contact frame (as $\Delta\boldsymbol{\xi}$ and $\mathbf{F}_{\text{t\_old}}$);
use the J\"{a}ger algorithm
to find $\mathbf{F}_{\text{t}}$ in the local contact frame; 
and then transform
the local $\mathbf{F}_{\text{t}}$ back into the global
$\overline{\mathbf{F}}_{\text{t}}$ so that this
force can be used to advance the particles' positions and orientations.
The global frame (overbar vectors with three components) 
is used throughout the
DEM code, except within the J\"{a}ger algorithm, where the local contact frame
is in effect (i.e., the bare vectors with two possibly non-zero components:
$\Delta\xi_{1}$-$\Delta\xi_{2}$ 
and $F_{\text{t},1}$-$F_{\text{t},2}$).
During these coordinate transformations, the $\mathbf{p}^{\ast}$ list can remain
within the contact frame, since it is not needed outside of the
J\"{a}ger algorithm.
\par
The incremental rotation of the local frame, $\Delta\overline{\boldsymbol{\omega}}$, is the sum
of two types of rigid rotation:
\begin{equation}
\Delta\overline{\boldsymbol{\omega}} = \Delta\overline{\boldsymbol{\omega}}^{\text{twirl}} \:+\: \overline{\mathbf{n}}^{p} \times \Delta\overline{\mathbf{n}}^{p}
\label{eq:dOmegab}
\end{equation}
a rigid twirling of the two particles about axis $\overline{\mathbf{n}}^{p}$,
\begin{equation}
\Delta\overline{\boldsymbol{\omega}}^{\text{twirl}} = 
   \frac{1}{2}\left[ \left( \dTb{p} + \dTb{q} \right)\boldsymbol{\cdot}\overline{\mathbf{n}}^{p} \right] \overline{\mathbf{n}}^{p}
\label{eq:dTwirl}
\end{equation}
and a rigid tilting $\Delta\overline{\mathbf{n}}^{p}$ of the normal vector,
\begin{equation}
\Delta\overline{\mathbf{n}}^{p} = \dTb{p}\times\overline{\mathbf{n}}^{p} \:+\: 
  \Kb{p}\boldsymbol{\cdot}\left( \Kb{p} + \Kb{q}  \right)^{-1}\!\!\boldsymbol{\cdot}
  \left[ ( \dTb{q} - \dTb{p} )\times\overline{\mathbf{n}}^{p} - 2\Kb{q}\boldsymbol{\cdot}\Delta\overline{\boldsymbol{\xi}} \right]
\label{eq:dNp}
\end{equation}
The latter expression involves the curvature tensors
$\Kb{p}$ and $\Kb{q}$ as described in~\cite{Kuhn:2004b}.
For the simple case of two spherical surfaces,
\begin{equation}
\Delta\overline{\mathbf{n}}^{p} = \dTb{p}\times\overline{\mathbf{n}}^{p} \:+\: 
  \frac{R^{q}}{R^{p}+R^{q}} 
  \left[ ( \dTb{q} - \dTb{p} )\times\overline{\mathbf{n}}^{p} + \frac{2}{R^{q}} \Delta\overline{\boldsymbol{\xi}} \right]
\label{eq:dNpSimple}
\end{equation}
where $R^{p}$ and $R^{q}$ are the surface radii of curvature.
\par
Quaternions offer a compact and efficient means of shifting between
coordinate frames~\cite{Altmann:1986a}.
They are commonly used in DEM codes to store particle orientations
and effect particle rotations~\cite{Vuquoc:2000a}.
The orientation of the contact plane, however, is distinct from
the orientations of the two particles, and
we will use quaternions to rotate the contact plane and transform
between the 
$\Delta\overline{\boldsymbol{\xi}}$--$\overline{\mathbf{F}}_{\text{t}}$ 
and $\Delta\boldsymbol{\xi}$--$\mathbf{F}_{\text{t}}$ pairs.
\par
The frame orientation of a contact is expressed as the four-component
unit quaternion $\overset{\circ}{\mathbf{q}} = \langle q_{0}, \mathbf{q} \rangle = 
\langle q_{0}, q_{1}, q_{2},q_{3}\rangle$ having a unit norm,
$|\overset{\circ}{\mathbf{q}} |^{2} = \sum{q_{i}^{2}} = 1$.
When two particles first touch, their contact's quaternion 
is initialized as, say,
\begin{equation}
\overset{\circ}{\mathbf{q}} = \left\langle\,
      \cos(\beta/2),\,\sin(\beta/2)[\sin(\eta),\,-\cos(\eta),\,0]
      \,\right\rangle
\label{eq:qinit}
\end{equation}
which is formed from two Euler angles of the contact frame:
$\beta=\cos^{-1}(\overline{n}^{p}_{3})$ and
$\eta=\tan^{-1}(\overline{n}^{p}_{2} / \overline{n}^{p}_{1})$.
With every time step and before entering the J\"{a}ger algorithm, the $\overset{\circ}{\mathbf{q}}$ of each contact is updated to account
for the contact's incremental rotation 
$\Delta\overline{\boldsymbol{\omega}}$ during the previous step (computed with
equation~\ref{eq:dOmegab}) which is embedded in a quaternion
$\Delta\overset{\circ}{\boldsymbol{\omega}}=\langle 0, \Delta\overline{\boldsymbol{\omega}} \rangle$:
\begin{equation}
\overset{\circ}{\mathbf{q}} \approx \overset{\circ}{\mathbf{q}}_{\text{old}} + \frac{1}{2}\Delta\overset{\circ}{\boldsymbol{\omega}}\circ\overset{\circ}{\mathbf{q}}_{\text{old}}
\label{eq:dqo}
\end{equation}
using the quaternion product
$\overset{\circ}{\mathbf{a}}\circ\overset{\circ}{\mathbf{b}}$
that is defined in the Appendix.
Equation~(\ref{eq:dqo}) is an approximation that can cause $\overset{\circ}{\mathbf{q}}$ to
drift slightly from the unit condition $|\overset{\circ}{\mathbf{q}}|=1$.
If desired, $\overset{\circ}{\mathbf{q}}$ can be renormalized, either exactly as
$\overset{\circ}{\mathbf{q}} 
\leftarrow \overset{\circ}{\mathbf{q}}/|\overset{\circ}{\mathbf{q}}|$ 
or with the Katz approximation~\cite{Katz:1997a}.
Before entering the J\"{a}ger algorithm, the contact's tangential
movement $\Delta\overline{\boldsymbol{\xi}}$ and previous force $\overline{\mathbf{F}}_{\text{t\_old}}$ are rotated into the
contact frame,
\begin{align}
\Delta\boldsymbol{\xi} &= \overset{\circ}{\mathbf{q}} \circ \Delta\overline{\boldsymbol{\xi}} \circ \overset{\circ}{\mathbf{q}}\!\,^{\ast}     \label{eq:dXiq}\\
\mathbf{F}_{\text{t\_old}} &= \overset{\circ}{\mathbf{q}} \circ \overline{\mathbf{F}}_{\text{t\_old}} \circ \overset{\circ}{\mathbf{q}}\!\,^{\ast} \label{eq:dFtoldq}
\end{align}
using the quaternion
conjugate $\overset{\circ}{\mathbf{q}}\!\,^{\ast}$ and the double product defined in the Appendix.
After leaving the J\"{a}ger algorithm, the returned tangential force is 
rotated back into the global frame:
\begin{equation}
\overline{\mathbf{F}}_{\text{t\_old}} = \overset{\circ}{\mathbf{q}}\!\,^{\ast} \circ \mathbf{F}_{\text{t\_old}} \circ \overset{\circ}{\mathbf{q}}
\label{eq:dFtoldbq}
\end{equation}
where the force then would be included in Newton's equations to
advance the particles' positions and orientations.
\subsection{Data structures}\label{sec:lists}
\begin{figure}[tb]
\centering
\includegraphics[scale=0.9]{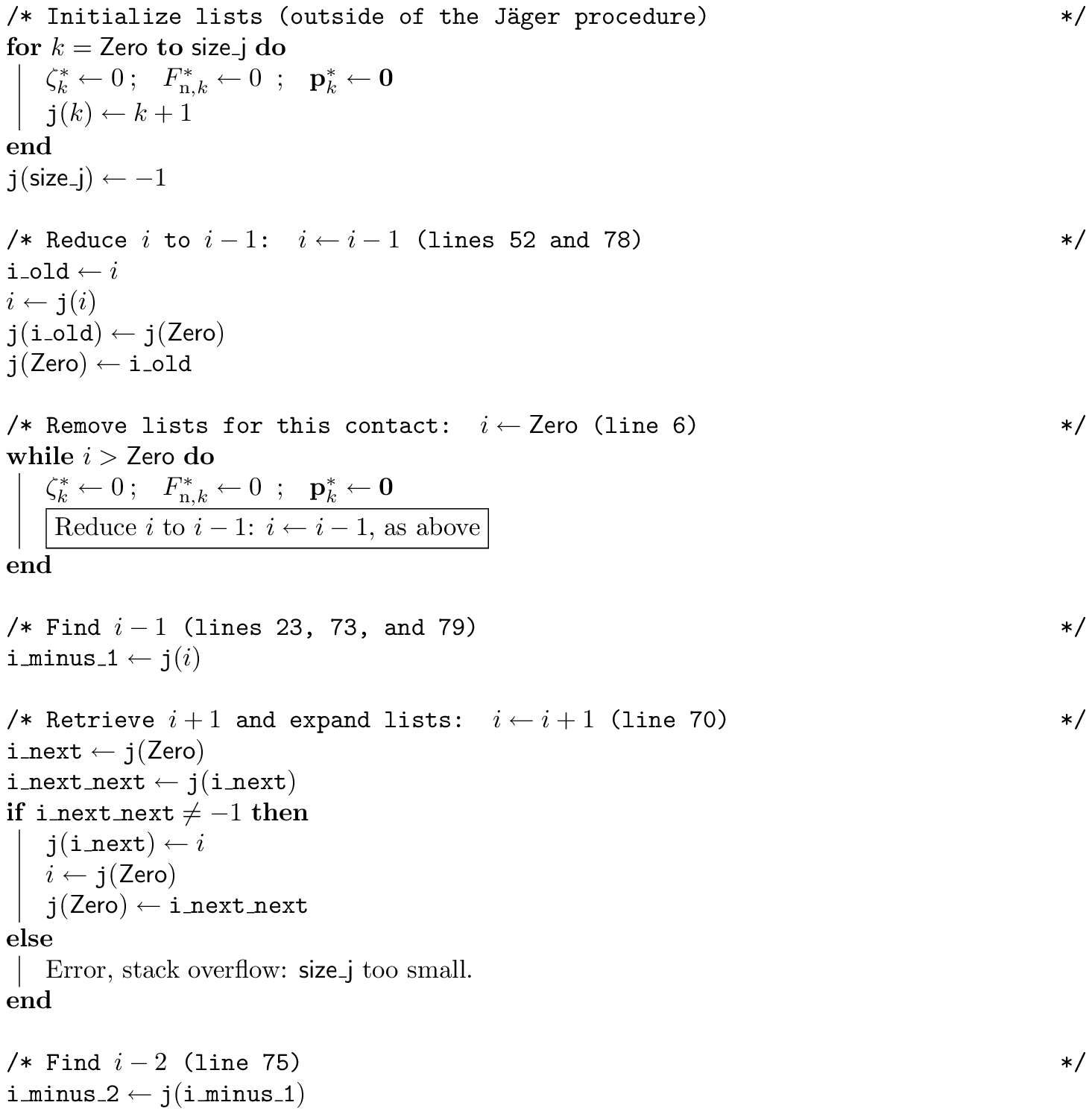}
\caption{\label{fig:Jager4}
         Maintaining linked lists of the equivalent load history.}
\end{figure}

The equivalent load history of each contact is stored in three lists
$\zeta^{\ast}_{0,1,\ldots,i}$, 
$F^{\ast}_{\text{n},0,1,\ldots,i}$, and
$\mathbf{p}^{\ast}_{0,1,\ldots,i}$.
These lists are stack structures with
last-in/first-out access and with no need for random access into the stacks.
Linked lists are compact means of storing and accessing data 
of this form (see~\cite{Knuth:1973a},~\S2.2.3).
Rather than creating separate arrays for each contact, 
three master-lists 
$\zeta^{\ast}$, $F_{\text{n}}^{\ast}$, and $\mathbf{p}^{\ast}$ 
can store the data of all contacts.
Figure~\ref{fig:Jager4} gives code for accessing and modifying
these lists at various
points within Figures~\ref{fig:Jager1}, \ref{fig:Jager2}, and~\ref{fig:Jager3}.
The input argument $i$ of the J\"{a}ger procedure 
(line~1 of Figure~\ref{fig:Jager1})
is a pointer to the top position of a
contact's history.
A fourth list $j$ stores descending pointers to the next (lower)
position of each contact's stack, and it also stores
ascending pointers to the next available position for placing new data.
In this sense, $\Zetasi{i}$ or $\zeta^{\ast}$\texttt{(i)} 
is the top of the $\zeta^{\ast}$-stack of a particular contact;
$\zeta^{\ast}$\texttt{(j(i))} is the next lower position,
$\zeta^{\ast}_{i-1}$, in its stack; etc.
The \texttt{j(Zero)} value points to the next available position;
\texttt{j(j(Zero))} points to the following available position; etc.
Pointer $i$ is associated with a single contact, and it
can also point to other data (lists) for the contact:
in particular, its $\overset{\circ}{\mathbf{q}}$ and $\mathbf{F}_{\text{t\_old}}$ data.
These lists have a length as long as the number of contacts.
The pointer list $j$, however, is associated with the J\"{a}ger equivalent load
histories of all contacts, and the four lists $j$, $\zeta^{\ast}$, $F_{\text{n}}^{\ast}$, and $\mathbf{p}^{\ast}$
are as long as the number of contacts times the average length
of the load history per contact:
the capacity parameter \textsf{size\_j} in Figure~\ref{fig:Jager4}.
\section{ALGORITHM PERFORMANCE}\label{sec:perform}
\subsection{Performance in a simple loading-unloading 
            sequence}\label{section:compare}
\begin{figure}
\centering
\includegraphics{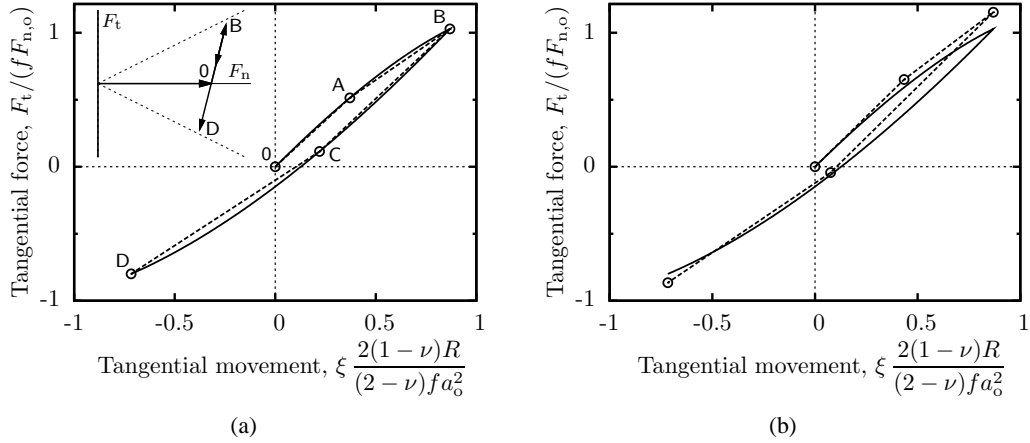}
\caption{\label{fig:Compare}
 Comparisons of the Mindlin-Deresiewicz solution 
 with (a) the J\"{a}ger algorithm and (b) an
 incremental-stiffness algorithm~\cite{Zhang:2007a}.  
 The proportional loading-unloading sequence is shown in the inset.
}
\end{figure}

In section~17 of~\cite{Mindlin:1953a}, Mindlin and Deresiewicz
give a closed-form solution of movement and tangential force for
a simple sequence of proportional loading and unloading.
This sequence will serve to illustrate the accuracy of the
J\"{a}ger algorithm and to compare it with an
incremental-stiffness solution~--- the displacement-driven
algorithm of Zhang and Vu-Quoc~\cite{Zhang:2007a},
employed as an elastic-frictional formulation.
In the first stage of loading, two elastic spheres are pressed together with
force $F_{\text{n},0}$ (see inset of Figure~\ref{fig:Compare}a).
The tangential and normal forces are then increased
proportionally, with $dF_{\text{t}}/dF_{\text{n}}=4>f$, until
$F_{\text{t}}$ nears the frictional limit, 
$F_{\text{t,B}} = 0.9 f F_{\text{n,B}}$.
The tangential and normal forces are then reversed in the same proportion
until $F_{\text{t,D}} = -0.9 f F_{\text{n,D}}$.
The solid lines in Figure~\ref{fig:Compare} are the Mindlin-Deresiewicz
solution (\cite{Mindlin:1953a}, equations~80, 83, and~84).
This closed-form solution is faithfully reproduced by both
the J\"{a}ger method and
by the incremental-stiffness method, provided that
the displacement steps ($\Delta\zeta$,$\Delta\xi$) are infinitesimal.
A more telling test is when the loading sequence is broken into four
large steps, with two displacement steps between 
$F_{\text{t,0}}$ and $F_{\text{t,B}}$, and two
more steps between $F_{\text{t,B}}$ and $F_{\text{t,D}}$.
Although these steps are much larger than would be expected
in DEM simulations,
the J\"{a}ger method gives the exact Mindlin-Deresiewicz solution
for each large loading step (Figure~\ref{fig:Compare}a);
indeed, when the J\"{a}ger solution is expanded, it coincides with
the closed-form Mindlin-Deresiewicz solution.
The incremental-stiffness solution is seen to yield an
approximation of the Mindlin-Deresiewicz solution
(Figure~\ref{fig:Compare}b), 
although this approximation can be greatly improved
by sub-dividing each large step into smaller sub-increments.
\par
Even though shearing tractions within the contact area
are rarely computed in DEM simulations, a simple change to the
J\"{a}ger algorithm yields the shearing traction $\boldsymbol{\tau}$
in addition to the tangential force $\mathbf{F}_{\text{t}}$.
We now consider such tractions in order to illustrate the exact
correspondence of the J\"{a}ger and Mindlin-Deresiewicz
solutions for this simple example of proportional loading and unloading
(\cite{Mindlin:1953a}, \S14, 15, and~16).
Whenever a tangential force
($\mathbf{F}_{\text{t}}$, $\mathbf{F}_{\text{t\_old}}$,
$\Delta\mathbf{F}_{\text{t}}$, etc.)
appears within a line in Figures~\ref{fig:Jager1},
\ref{fig:Jager2}, and~\ref{fig:Jager3},
the corresponding tangential traction can be computed by 
simply replacing this force with a traction
distribution $\boldsymbol{\tau}(\rho )$ that is computed
by replacing, in turn, that line's normal forces
($\mathbf{F}_{\text{n}}$, $\mathbf{F}_{\text{n\_old}}$,
$\mathbf{F}_{\text{n}}^{\ast}$, etc.)
with Catteneo-Mindlin distributions,
\begin{equation}
\sigma_{\text{C-M}}(\rho) =
\begin{cases}\displaystyle
\frac{3\overline{E}}{2\pi R}\left( a^{2} - \rho^{2}\right)^{1/2}
  & \rho < a \\
0 & \rho \ge a
\end{cases}
\label{eq:CM}
\end{equation}
In this distribution, $\rho$ is the radial distance from the center of the 
circular contact area, and ``$a$'' is a radius computed with
equation~(\ref{eq:a}) using the corresponding normal force
($\mathbf{F}_{\text{n}}$, $\mathbf{F}_{\text{n\_old}}$,
$\mathbf{F}_{\text{n}}^{\ast}$, etc.).
\begin{figure}[tb]
\centering
\includegraphics[scale=0.9]{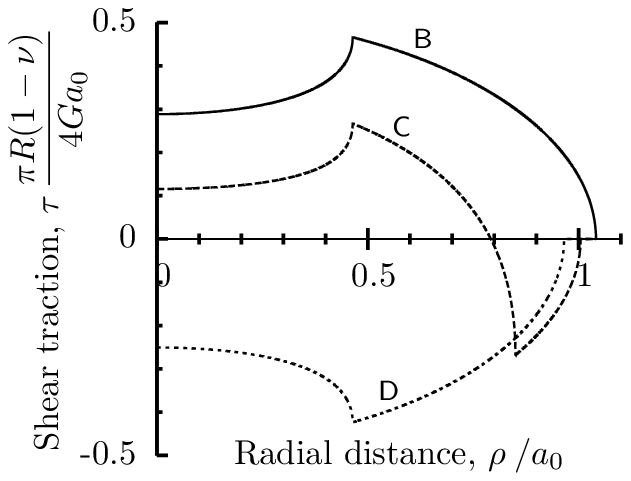}
\caption{\label{fig:TractionCompare}
         Shear tractions from both the 
         J\"{a}ger and Mindlin-Deresiewicz solutions
         at three stages of a
         loading and unloading sequence (see Figure~\ref{fig:Compare}a).}
\end{figure}
Figure~\ref{fig:TractionCompare} shows the J\"{a}ger 
and Mindlin-Deresiewicz solutions at three stages in the 
loading and unloading sequence of Figure~\ref{fig:Compare}a.
The two solutions coincide exactly.
\subsection{Performance with a large assembly of spheres}
The algorithm's computational speed and memory demand are examined with
a DEM simulation of biaxial plane-strain compression of an assembly of
4096 spheres.  
The densely packed assembly was compressed in the vertical direction while maintaining constant stress in one horizontal direction and zero strain
in the other horizontal direction.
As is typical in such tests, the initial deformation is primarily
elastic but is followed by inelastic behavior and intense dilation,
with the assembly reaching a peak compressive stress at a strain of
about $2.5\%$~\cite{Kuhn:2010a}.
The simulation produced vertical strains of 0--4\% (40,000 time steps)
during which the assembly contained an average of about 8600 contacts.
The two simulations were run on a single-core single-thread
Pentium-4 2.66GHz processor with 333MHz memory.
\par
Standard DEM simulations use a central difference explicit time
integration scheme in which numerical stability can usually be assured
when the time step is less than some limiting value.
Tavarez and Plesha~\cite{Tavarez:2007a} give the maximum time
step as $2(\sqrt{1-\xi} - \xi)/\omega_{\text{max}}$,
where $\xi$ is the damping expressed as a fraction of
critical damping, and $\omega_{\text{max}}$ is the highest
vibrational frequency of the entire assembly.
Because $\omega_{\text{max}}$ is difficult to assess,
most simulations use a simplified approach based on the contact stiffnesses
and particle masses, in which the time
step is less than some fraction of $\sqrt{m/k}$
(e.g.,~\cite{OSullivan:2004a}).
With Hertz contacts, stiffness depends on the contact force,
and the time step must be adjusted accordingly.
The tangential stiffness of the Cattaneo-Mindlin-Deresiewicz contact
is no greater than the normal (Hertz) stiffness, and the tangential
behavior softens with increasing obliquity of the 
contact force.
For these reasons, an upper bound of the time step
can be based upon the normal forces and the corresponding Hertz
stiffnesses (see~\cite{Vuquoc:2000a,Ashmawy:2003a,Uthus:2008a}
for instructions on adaptive time steps with Hertz contacts).
\par
Table~\ref{table:performance} compares the computation time for 
a simulation that employed the 
J\"{a}ger algorithm with one that used a simpler incremental-stiffness
modified-Mindlin model, a model that
can capture only a single force reversal (e.g.,~\cite{Lin:1995a}).
%
%
%
%
A run-time profiler (\textsf{gprof}) was used to determine the CPU-time
spent computing the contact forces within each simulation.
The CPU-time spent within the J\"{a}ger algorithm was considerably
greater that within the simpler procedure (by a factor of $2.4$).
A DEM simulation requires other calculations besides 
force computation,
and the total time to run the simulations
was only modestly longer when using the J\"{a}ger 
algorithm (by a factor of $1.5$).
\par
The maximum memory required by the J\"{a}ger algorithm occurred at a strain
of $0.5$\%, when the assembly contained about 8700 contacts~--- during 
a period of substantial inelastic loading but prior to the peak stress.
The corresponding average length of the equivalent load history per contact
is also shown in Table~\ref{table:performance}.
The lists had average lengths as great as 260 per contact with the J\"{a}ger
algorithm.
Using four 8-byte floating-point numbers for $\zeta^{\ast}$,
$F_{\text{n}}^{\ast}$, and $\mathbf{p}^{\ast}$ and a 4-byte integer
for $j$, an average history of length 260 requires 9630~bytes
of memory per contact~--- about 83~MBytes for the 8700 contacts.
Without this demand, the entire DEM code would require only 10~MBytes
of memory, so the J\"{a}ger algorithm does impose a substantial memory
demand on the simulations.
By using the approximation described in Section~\ref{sec:approx},
however,
the required memory can be reduced considerably.
With a parameter $|\Delta\mathbf{p}_{\text{max}}|$ of 0.10$f$,
which would produce only small errors in the force calculations, the equivalent
history is reduced in length from 260 to 171; a larger parameter of 0.20$f$
reduces the average length to 107 (about 40\% of the length 260
required for exact force calculations).
\section{CONCLUSION}
The J\"{a}ger algorithm is an efficient approach to computing the
three-dimensional
Cattaneo-Mindlin-Deresiewicz contact force for arbitrary contact
translations, regardless of size or sequence.
Each movement episode is solved exactly and in whole, without the need
to divide the movement into smaller increments.
The algorithm is particularly useful for DEM simulations 
in which particles are persistently in non-simple contact
across many time steps, such as simulations of large inelastic deformation or
of cyclic loading. 
The paper details the necessities for incorporating 
the algorithm into DEM codes.
The computation time is only about 50\% greater than when a much simpler
contact algorithm is employed, 
a substantial increase but certainly not an obstacle to its adoption.
More problematic, however, is the considerable storage required by the
algorithm.
Based on the results of a fairly severe performance test, 1~GByte of memory
can accommodate DEM simulations of about 40,000 particles, 
and perhaps 100,000 
particles when an approximation is used within the algorithm.
Because it authentically captures the contact mechanics of 
ideal elastic particles, however, the algorithm should 
be considered for simulations that
aim for fidelity to the true behavior of such granular materials.
\par
\ack
This material is based upon work supported by the National Science Foundation under Grant No. NEESR-936408.
\section*{APPENDIX}
The quaternion product in (\ref{eq:dqo}) is the matrix
product
\begin{equation}
\frac{1}{2}\Delta\overset{\circ}{\boldsymbol{\omega}}\circ\overset{\circ}{\mathbf{q}}_{\text{old}} =
\frac{1}{2}
\begin{bmatrix}
         0 & -\dOmega{1} & -\dOmega{2} & -\dOmega{3} \\
\dOmega{1} &           0 & -\dOmega{3} &  \dOmega{2} \\
\dOmega{2} &  \dOmega{3} &           0 & -\dOmega{1} \\
\dOmega{3} & -\dOmega{2} &  \dOmega{1} & 0
\end{bmatrix}
\begin{bmatrix}
q_{\text{old},0} \\
q_{\text{old},1} \\
q_{\text{old},2} \\
q_{\text{old},3}
\end{bmatrix}
\end{equation}
The vector rotations on the right of 
Eqs.~(\ref{eq:dXiq}), (\ref{eq:dFtoldq}), and~(\ref{eq:dFtoldbq})
are the matrix products
\begin{align}
\overset{\circ}{\mathbf{q}} \circ \mathbf{a} \circ \overset{\circ}{\mathbf{q}}\!\,^{\ast} &= 2[\mathbf{Q}]^{\text{T}}[\mathbf{a}]\\
\overset{\circ}{\mathbf{q}}\!\,^{\ast} \circ \mathbf{a} \circ \overset{\circ}{\mathbf{q}} &= 2[\mathbf{Q}][\mathbf{a}]
\end{align}
where $\overset{\circ}{\mathbf{q}}$ is a unit quaternion, $\overset{\circ}{\mathbf{q}}\!\,^{\ast}$ is its conjugate,
$\mathbf{a}$ is a 3-vector, and
\begin{equation}
\left[ \mathbf{Q} \right] =
\begin{bmatrix}
\frac{1}{2}-q_{2}^{2}-q_{3}^{2} & 
                              q_{0}q_{3}+q_{1}q_{2} & q_{1}q_{3}-q_{0}q_{2}\\
q_{1}q_{2}-q_{0}q_{3} & \frac{1}{2}-q_{1}^{2}-q_{3}^{2} &
   q_{0}q_{1}+q_{2}q_{3} \\
q_{1}q_{3}+q_{0}q_{2} & q_{2}q_{3}-q_{0}q_{1} &
   \frac{1}{2}-q_{1}^{2}-q_{2}^{2}
\end{bmatrix}
\end{equation}
%
%
%

%
%
\pagebreak
\begin{table}
\figcap{5.00in}
\centering
\caption{\label{table:performance}Performance during a simulation
of biaxial plane-strain compression: the J\"{a}ger and conventional
modified-Mindlin algorithms.}
\begin{small}
\begin{tabular}{lcc}
\toprule
& \multicolumn{2}{c}{Algorithm}\\
\cline{2-3}
& modified-Mindlin & J\"{a}ger \\
\midrule
Force calculation time (minutes) & 3.4 & 8.2 \\
Total simulation time (minutes)  & 17.6 & 26.2 \\ 
\midrule
Avg. length of equiv. history, exact J\"{a}ger forces & -- & 260 \\
Avg. length of equiv. history, $|\Delta\mathbf{p}_{\text{max}}|/f = 0.10$
  & -- & 171 \\
Avg. length of equiv. history, $|\Delta\mathbf{p}_{\text{max}}|/f = 0.20$
  & -- & 107 \\
\bottomrule
\end{tabular}
\end{small}
\end{table}
\end{document}